\documentclass[
10pt,tightenlines, amsmath, amssymb, nofootinbib, prd,twocolumn,
superscriptaddress, showpacs, preprintnumbers]{revtex4}

\usepackage{graphicx}
\usepackage[figuresright]{rotating}

\newcommand{\beq}{\begin{equation}}
\newcommand{\eeq}{\end{equation}}
\newcommand{\bea}{\begin{eqnarray}}
\newcommand{\eea}{\end{eqnarray}}
\newcommand{\nn}{\nonumber}

\def\la{\langle }
\def\ra{ \rangle }

\newcommand{\MM}{{\cal M}}

\newcommand{\junk}[1]{}

\begin{document}

 \title{ Confinement- Deconfinement Phase Transition and Fractional Instanton Quarks 
 in Dense Matter} 
\author{Ariel R. Zhitnitsky\footnote{Invited talk delivered
  at the Light Cone Workshop,  July 7-15, 2005, Cairns, Australia.}}
\affiliation{Department of Physics and Astronomy, University of
  British Columbia, Vancouver,  Canada}

\begin{abstract}
We present arguments suggesting that  large size
overlapping instantons 
are the driving mechanism of the confinement-deconfinement  phase
transition at nonzero chemical potential $\mu$. 
The arguments are based on the picture that instantons at  very large 
chemical potential in the weak coupling regime are localized  
configurations with finite size $\rho\sim\mu^{-1}$. At the same time, 
 the same instantons at  smaller chemical potential
  in the strong  coupling regime are well represented
 by the so-called instanton-quarks with fractional topological charge
 $1/N_c$.  
We  estimate the critical chemical potential
$\mu_c(T)$ where    transition  between these two  regimes  takes place.
We identify this transition with  confinement- deconfinement phase transition.
    We also  argue that the instanton quarks
carry magnetic charges. As a consequence of it, there is a  relation
between our picture and the standard t'Hooft and Mandelstam  picture of the confinement.
We also comment on possible relations  of instanton-quarks with  ``periodic instantons"  ,  `` center vortices" ,   and ``fractional instantons"  in the brane construction.
We also argue that  the variation of 
the external parameter $\mu$, which plays the role of the vacuum expectation value
of a ``Higgs"  field at $\mu\gg \Lambda_{QCD}$,  allows to study the transition
 from a ``Higgs -like"  gauge theory   (weak coupling regime, $\mu\gg \Lambda_{QCD}$)
  to   ordinary QCD (strong coupling regime, $\mu\ll \Lambda_{QCD}$).  We also comment on some    recent lattice results on topological charge density distribution which
  support  our picture.
 
 \end{abstract}

\maketitle

\section{ Introduction}  

This talk is based
  on a number of original results\cite{TZ}-\cite{JZ} obtained  with different collaborators 
  at different times.
   
Color confinement, spontaneous breaking of chiral symmetry, 
the $U(1)$ problem and the $\theta$ dependence are some of the most
interesting questions in $QCD$. Unfortunately,  progress in the
understanding of these problems has been 
extremely slow. At the end of the 1970's A. M. Polyakov \cite{Po77}
demonstrated charge confinement in $QED_3$.  This was the first example
where nontrivial dynamics was shown to be a key ingredient for
confinement: 
The instantons  (the monopoles in 3d) play a crucial role
in  the dynamics of confinement in $QED_3$.   
 Instantons in four dimensional QCD were discovered  30 (!)  years ago
\cite{Belavin:1975fg}. However, their role in   $QCD_4$ remains
unclear even today due to the divergence  of the instanton density for 
large size instantons. 

Approximately at the same time instanton dynamics  was developed
in two dimensional, classically conformal, 
  asymptotically free models (which may have some analogies  with
  $QCD_4$). Namely, using an exact accounting and resummation
of the $n$-instanton solutions in $2d~CP^{N_c-1}$ models,
 the original problem of a statistical instanton
 ensemble  was   mapped unto a $2d$-Coulomb Gas (CG) 
system of pseudo-particles 
with fractional topological charges $\sim 1/N_c$ (the so-called
instanton-quarks) \cite{Fateev}.   The instanton-quarks do not exist
separately as individual objects. Rather, they appear in the system
all together as a set of $\sim N_c$ instanton-quarks so that the total
topological 
charge of each configuration is always an integer. This means that a
charge for an individual instanton-quark cannot be created
and measured.
Instead, only the total topological charge for the whole configuration
is forced to be integer and has a physical meaning. This picture leads to the 
elegant explanation of confinement and other important properties of
the $2d~CP^{N_c-1}$ models \cite{Fateev}.  
Unfortunately, despite some attempts \cite{Belavin}, there is no
demonstration that a
similar picture occurs in $4d$ gauge theories,  
where the instanton-quarks would become the relevant  quasiparticles. 
Nevertheless, there remains  a strong 
  suspicion that this picture, which assumes that instanton-quarks with
   fractional topological charges  $\sim 1/N_c$ become the relevant 
  degrees of freedom in the confined phase, may be correct in $QCD_4$.
 
 On the phenomenological side, the development of the 
instanton liquid model (ILM) \cite{ILM,shuryak_rev} has encountered 
  successes (chiral symmetry breaking, resolution of the $U(1)$
 problem, etc) and 
 failures (confinement could not be described by well separated and
 localized lumps with integer 
 topological charges). Therefore, it is fair to say that
 at present, the widely accepted viewpoint is that
 the  ILM can explain  many experimental data
  (such as hadron masses, widths, correlation functions, decay
  couplings, etc), with one, but crucial exception: confinement.
  There are many arguments against the ILM approach, see e.g. \cite{Horvath}, 
  there are many arguments supporting it \cite{shuryak_rev}.

  In this talk we present  new arguments supporting the idea
  that the instanton-quarks along with instantons 
  are the relevant quasiparticles in the strong coupling regime.
   In this case, 
  many problems
formulated in \cite{Horvath} are naturally resolved as both phenomena,
confinement and chiral symmetry breaking are originated from the same
vacuum configurations, instantons,  
which may have arbitrary scales: the finite size localized lumps with integer topological
charges, as well as 
 set of $N_c$ fractionally  $1/N_c$ -charged    correlated objects sitting at 
 arbitrary large distances from each other. 
In this picture when    fractionally  charged  $1/N_c$ constituents   
   propagate far away from each other, the confinement could be  a natural consequence
   of a dynamics of these well correlated objects.
  We emphasize that 
 along with instanton quarks there are ordinary  instantons with integer topological
 charges. Indeed, if the instanton-quarks 
  are close to each other they bound together and 
  likely to form an ordinary instanton.  If the instanton quarks
  far away from each other, the description in terms of fundamental instanton quarks
  is more appropriate.
  The precise probability  
  for each configuration depends on the interplay between action and entropy. 
   Such a  feature when the well- localized instantons and de-localized instanton- quarks coexist 
   may lead to the understanding why the  chiral
  symmetry breaking phenomenon ( which, as ILM suggests \cite{shuryak_rev}, is 
  due to the  well- localized instantons)
  and  the confinement - deconfinement phase transitions (which is due to,
   the de-localized instanton quarks, according to the present proposal) are so close
  to each other. In our picture such  a ``conspiracy"  is
   a  simple reflection of  the fact that both phenomena
  are due to the same configurations, instantons, which however can be in different configurations.

More importantly, 
  we make some very specific predictions which can be tested with 
  traditional Monte Carlo techniques, by studying QCD at nonzero isospin
  chemical potential\cite{Isospin}.  
  
  We start in  Section II  by reviewing  recent work for QCD 
at large $\mu$ in the deconfined phase \cite{ssz}, where the
  instanton calculations are under complete  
  theoretical control, since the instantons are
  well-localized objects with a typical size 
  $\rho\sim 1/\mu$.
  
We then discuss in Section III the dual representation of the
low-energy effective chiral Lagrangian in the regime
of small chemical potential where confinement takes place.
 We shall argue that the corresponding dual representation
 corresponds to a statistical 
system of interacting pseudo-particles with fractional $1/N_c$ topological
charges  which can be  identified   with instanton-quarks \cite{JZ} 
suspected long ago \cite{Fateev,Belavin}.

Based on these observations  we   make a {\bf conjecture}\cite{TZ}  formulated 
in Section  IV  that the transition from 
the description in terms of well localized instantons with finite size
at large $\mu$  to the description 
in terms of the instanton quarks with fractional $1/N_c$ topological
charges precisely corresponds to the deconfinement-confinement phase
transition.

In Section V  we explicitly calculate the critical chemical
potential $\mu_c$ where this phase transition should occur. Our conjecture
can be explicitly and readily tested in numerical simulations
due to the absence  of the  sign problem  at arbitrary value of the 
isospin chemical potential. If our conjecture turns out to be
correct, it would be   
an explicit  demonstration of the link between confinement and instantons.

In section VI   we present some arguments explaining why the standard 
picture of confinement suggested long ago by t'Hooft and Mandelstam\cite{Hooft}
( which is based on  the condensation of the magnetic monopoles)
is consistent with  our interpretation of the confinement  when  the instanton quarks
play the key role.
Finally,  Section VII is our conclusion where we argue that our picture of the confinement deconfinement phase transition  can be tested on the lattice  with 
  traditional Monte Carlo techniques if one studies  QCD at nonzero isospin (rather than baryon) 
  chemical potential.  We also comment on relations with different works.
  Finally, we make some comments on   recent lattice results on topological 
  density distribution.

 \section{ Instantons at large $\mu$}
At low energy and large chemical potential, the $\eta'$ is light and 
described by the Lagrangian derived in \cite{ssz}:
\begin{eqnarray}
  \label{Leta}
  L_\varphi=f^2(\mu) [(\partial_0\varphi)^2-u^2 (\partial_i\varphi)^2] -
  V_{inst}(\varphi). 
\end{eqnarray}
where the $\varphi$ decay constant, $f^2(\mu_B)\!=\!\mu_B^2/8\pi^2$ and
$f^2(\mu_I)\!=\!3\mu_I^2/16\pi^2$, and its velocity,
$u^2\!=\!1/3$ \cite{ssz,schaefer}. 
We define baryon and isospin chemical potentials as
$\mu_{B,I}\!=\!(\mu_u\pm\mu_d)/2$.    
The nonperturbative potential $V_{inst}\sim \cos(\varphi-\theta)$
is due to instantons, which are suppressed at large chemical potential.
  
The instanton-induced effective four-fermion interaction for 2
flavors, $u, d$, is given by
\cite{tHooft,SVZ},
\begin{eqnarray}
   L_{\rm inst}&=& \int\!d\rho\, n(\rho) 
  \biggl({4\over3}\pi^2\rho^3\biggr)^2 \biggl\{
  (\bar u_R u_L)(\bar d_R d_L) + \nonumber \\
  &+& {3\over32} \biggl[ (\bar u_R\lambda^a u_L)(\bar d_R\lambda^a d_L)\\
  &-& {3\over4}(\bar u_R\sigma_{\mu\nu}\lambda^a u_L)
    (\bar d_R\sigma_{\mu\nu}\lambda^a d_L) \biggr]
  \biggr\}  + {\rm H.c.}\nonumber
  \label{inst_vertex}
\end{eqnarray}
We study this problem at nonzero temperature and chemical
potential for $T\ll\mu$, and we use the standard formula
for the instanton density at two-loop order \cite{shuryak_rev}
\begin{eqnarray}
\label{instanton}
 n(\rho)&=& C_N(\beta_I(\rho))^{2N_c} \rho^{-5}
 \exp[-\beta_{II}(\rho)]  \\
 &\times&  \exp[-(N_f \mu^2 + \frac13 (2N_c+N_f) \pi^2
 T^2)\rho^2], \nn 
\end{eqnarray}
where
\begin{eqnarray}
  C_N &=& \frac{0.466 e^{-1.679N_c} 1.34^{N_f}}{(N_c-1)!(N_c-2)! },\nn\\
\beta_I(\rho)&=&-b \log(\rho\Lambda),\nn\\
\beta_{II}(\rho)&=&\beta_I(\rho)+\frac{b'}{2b} \log\left(\frac{2
  \beta_I(\rho)}{b}\right), \nn \\ 
b&=& \frac{11}3 N_c-\frac23 N_f,\nn\\
 b'&=&\frac{34}3 N_c^2-\frac{13}3 N_f
N_c +\frac{N_f}{N_c}.\nn
\end{eqnarray}
By taking the average of Eq.\ (\ref{inst_vertex}) over the
 state  with nonzero vacuum expectation value for the
 condensate,  one finds  
\bea
\label{Vinst2}
  V_{\rm inst}(\varphi)&=& -\int\!d\rho\, n(\rho)
  \biggl({4\over3}\pi^2\rho^3\biggr)^2 \nn\\
  &\times& 12|X(\mu)|^2\cos(\varphi-\theta) \\ 
 &=& -a(\mu,T)\mu^2\Delta^2\cos(\varphi-\theta),\nn
\eea 
where  $|X(\mu_B)|\!=\!3\mu_B^2\Delta/\sqrt{\beta_I(\rho)}$ and
$|X(\mu_I)|\!=\!3\sqrt{3}\mu_I^2\Delta/\sqrt{\beta_I(\rho)}$, and
$\Delta$ is the gap \cite{ssz,schaefer}.  
Therefore the mass of the $\varphi$ field is given by
\beq
  \label{massEta}
  m=\sqrt{\frac{a(\mu,T)}2} \frac{\mu \Delta}{f(\mu)}.
\eeq
The approach presented above is valid as long as the $\varphi$ field is
lighter than $\sim 2\Delta$, the mass of the other mesons in the
system \cite{ssz}, that is  if
\begin{eqnarray}
  \label{critical}
  a(\mu,T) \leq 8 f^2(\mu)/\mu^2. 
\end{eqnarray}
This is exactly the vicinity where the Debye screening scale and the
inverse gap become of the same order of magnitude \cite{ssz}, and
therefore, where the instanton expansion breaks down.

For reasons which will be clear soon, we want to represent 
 the  Sine-Gordon (SG) partition function (\ref{Leta}, \ref{Vinst2}) in
the equivalent dual Coulomb Gas (CG) representation \cite{ssz},
\bea
\label{CG}
Z =  \sum_{M_\pm=0}^\infty \frac{(\lambda/2)^M}{M_+!M_-!}  
\int d^4x_1 \ldots \int d^4x_M  \\
e^{-i\theta\sum_{a=0}^M Q_a}\cdot e^{-{1\over 2f^2u}\sum_{a>b=0}^M  Q_aQ_b
G(x_a-x_b)} ,\nn\\
G(x_a - x_b) = {1\over 4\pi^2 (x_a-x_b)^2}, ~~
\lambda \equiv {a\mu^2\Delta^2\over u}. \nn
\eea
Physical interpretation of the dual CG representation (\ref{CG}):\\
{\bf a) } Since
$Q_{\rm net}\equiv \sum_a Q_a$ is the total charge and it appears in
the action
multiplied be the parameter $\theta$, one
concludes that $Q_{\rm net}$ {\it is} the total topological
charge  of a given
configuration.\\
{\bf b)} Each charge 
$Q_a$ in a 
given configuration should be identified with an
  integer topological charge   well localized at the point $x_a$. This, 
by definition, 
corresponds to a small instanton positioned at $x_a$.\\
{\bf c)} While the starting low-energy effective Lagrangian contains only 
a   colorless field $\varphi$  we
have ended up with a representation of the partition function in which 
  objects carrying color (the instantons)   can be studied.\\
{\bf d)}
In particular, $II$ and $I\bar I$
interactions (at very  large distances) are exactly the same up to a
sign, order $g^0$, 
and are Coulomb-like.  This is in  contrast with semiclassical expressions
when  $II$ interaction is zero and $I\bar I$  interaction is order $1/g^2$.\\
{\bf e)} The very complicated picture of the    bare  $II$ and $I\bar I$
interactions 
becomes very simple for  dressed   instantons/anti-instantons
when all integrations over all  possible sizes, color orientations
and interactions with background fields are properly accounted for.\\
{\bf f)} As expected, the ensemble of small $\rho\sim 1/\mu$ instantons can
not produce confinement. 
 This is in accordance with the fact that there is no confinement at
 large $\mu$. 
 
\section{ Instantons at small $\mu$} 
 We   want to repeat the same procedure that led to the CG representation
 in the confined phase at small $\mu$ to see if any traces from the
 instantons  
 can be recovered. We start from the chiral Lagrangian and keep only
 the diagonal   
 elements of the chiral  matrix 
 $U=\exp\{i{\rm
diag}(\phi_1,\dots,\phi_{N_f})\}$ which are relevant in the
description of the ground state. 
 Singlet  combination is defined as  $\phi =
{\rm Tr}~U$.
  The effective Lagrangian for  the $\phi$  is
 \bea
 \label{chiral}
 L_{\phi}&=& f^2 
( \partial_{\mu} \phi)^2 + E \cos\left( \frac{ \phi - \theta }{N_c}\right) \nn\\
&+& \sum_{a=1}^{N_f} m_a \cos \phi_a
\eea
  A    Sine-Gordon structure for the singlet combination  
  corresponds to the following behavior of the $(2k)^{\rm th}$
   derivative of the vacuum energy in pure gluodynamics \cite{veneziano},
\bea
\frac{ \partial^{2k} E_{vac}(\theta)}{ \partial \, \theta^{2k}}|_{\theta=0} \sim \int \prod_{i=1}^{2k} dx_i \langle
Q(x_1)...Q(x_{2k})\rangle \nn\\
\sim (\frac{i}{N_c})^{2k}, ~~~~~~~ {\rm where}~~
Q\equiv\frac{g^2}{32\pi^2} G_{\mu\nu} {\widetilde G}_{\mu\nu}.
\eea
The same structure was also advocated in \cite{HZ} from a different
perspective. 
 As in (\ref{CG}) the Sine-Gordon effective field
 theory (\ref{chiral}) can be represented in terms 
 of a classical statistical ensemble (CG representation) similar to
 (\ref{CG})  with the replacements $\lambda\rightarrow E, ~ u\rightarrow 1$,
 more precisely,
 \bea
\label{CG1}
Z =  
\sum_{M =0}^\infty \frac{(E/2)^M}{M! }  \int d^4x_1 \dots \int d^4x_M \times\nn\\
\sum_{Q_a =\pm 1/N_c}\int D\phi e^{-f^2\int d^4x( \partial_{\mu} \phi)^2}\times \nn\\
 \left(e^{i \sum_{a=1}^M Q_a\left[\phi(x_a)-\theta\right] }\right) . 
 \eea
The functional integral is trivial to perform and one arrives at the dual CG action,
\bea
\label{CG2}
Z=\sum_{M_\pm=0}^\infty \frac{(E/2)^M}{M_+!M_-!}  
\int d^4x_1 \ldots \int d^4x_M\times  \nn\\
e^{-i\theta\sum_{a=0}^M Q_a }\cdot e^{-{1\over 2f^2}\sum_{a>b=0}  Q_aQ_b
G(x_a-x_b)} ,\nn\\
G(x_a - x_b) = {1\over 4\pi^2 (x_a-x_b)^2}. 
\eea
 The fundamental difference in comparison with the previous case
 (\ref{CG}) is that
while the total charge is integer, the individual charges are
   { \bf fractional $\pm 1/N_c$}.
This  is a
direct   consequence of the $\theta/N_c$ dependence in the underlying
effective Lagrangian (\ref{chiral}) before integrating out $\phi$ fields, see eq. (\ref{CG1}).\\
 Physical Interpretation of the  CG representation (\ref{CG2}) of
 theory (\ref{chiral}):\\  
 {\bf a)} As before,  one can identify
$Q_{\rm net}\equiv \sum_a Q_a$ with  the   total topological charge of
the given configuration.\\ 
{\bf b)} Due to the $2\pi$
periodicity of the theory, only configurations which contain an integer
topological number contribute to the partition function. Therefore, 
the number of particles for each given configuration $Q_i$ with
charges $ \sim 1/N_c$  
 must be proportional to $N_c$.\\
{\bf c) } Therefore, the number of integrations
over $d^4x_i$   in CS representation exactly equals $4 N_c k$, where 
$k$ is integer.   This number  $4 N_c k$ exactly
corresponds to the number of zero modes in the  $k$-instanton
background. This is basis for the conjecture \cite{JZ}  that 
at low energies (large distances) the fractionally charged species,
$Q_i=\pm 1/N_c$   
are the {\bf   instanton-quarks} suspected long ago \cite{Fateev}.\\
{\bf d)} For  the gauge group, $G$
the number of integrations would be equal to  $4k C_2(G)$  where
$C_2(G)$ is the quadratic Casimir of the gauge group
  ($\theta$ dependence in physical observables comes  in
the combination $\frac{\theta}{C_2(G)}$).   This number  $4k C_2(G)$
exactly corresponds 
to the  number of zero modes  in the $k$-instanton background for gauge
group $G$.\\
{\bf e)} The CG representation  corresponding to eq.(\ref{chiral}) describes
the confinement  phase of the theory.

One obvious  objection for such an identification of $Q_a$ with the
topological charge
immediately comes in mind: 
  it has long been known  that   instantons 
can explain most low energy QCD phenomenology \cite{ILM}   with the
 exception   confinement;  and we claim
that   confinement also arises in this picture: how can this be consistent?   
 We note that quark confinement can not be described in the  dilute gas 
approximation, when the instantons and anti-instantons are well
separated and maintain their individual properties (sizes, positions,
orientations), as it happens at large $\mu$.  However,   
in strongly coupled theories the instantons and
anti-instantons lose their individual properties (instantons will
``melt'')  their sizes become very large and they overlap.  The relevant
description is that of instanton-quarks  which can be far away from
each other,  but still  strongly correlated. For such configurations the confinement
is a possible outcome of the dynamics.

We should remark here that a precise 
form of the potential (\ref{chiral}) in the form of a single function $\sim \cos(\theta/N_c)$
is not a crucial issue for discussions below. A  combination of a number of terms,
$a_k\cos(k\theta/N_c)$
may change the interactions of the instanton
 quarks\footnote{We refer to ref. \cite{Diakonov:1999ae} 
where it is argued, based on  analysis of  two dimensional $CP^{N-1}$ model,
    that much more complicated
structure for the instanton quark interactions could result.}.  However, 
the most important element  here   remains the same:  the  
$\theta/N_c$ behavior   is well established result and remains untouched
even when more complicated terms are introduced. It will lead to the fractional charges 
$Q_a=\pm 1/N_c$ in Coulomb Gas representation (\ref{CG1},\ref{CG2}) in big contrast 
with weakly interacting phase  at large $\mu$ (\ref{CG}) where only integer topological charges appear.

We should also comment 
at this point that
our numerical estimates below are based exclusively on the instanton density at large $\mu$ 
while  we approaching the critical value. In this region the potential is well established and unique (\ref{Vinst2}). 
 Therefore, 
our results below  are not sensitive to the specific details of the potential (\ref{chiral}) at small $\mu$
when some  additional terms  might be present. 
 
\section{Conjecture.} 
We thus conjecture that the confinement-deconfinement phase transition
takes place at precisely the value where the dilute instanton
calculation breaks down.  At large $\mu$ the  weakly interacting phase (CS) is realized.
 Instantons are   well localized configurations with a typical size $\mu^{-1}$. Color in CS phase
  is not confined.     At low $\mu$ the strong interacting regime is realized and 
  color is confined. Instantons are not well localized configurations, but rather are represented by
  $N_c$ instanton quarks which can propagate far away from each other. The value of
the critical chemical potential as a function of temperature,
$\mu_c(T)$ is given by saturating the inequality (\ref{critical}).

Few remarks are in order. First, we can estimate the critical $\mu_c$ not only at $T=0$ , but also
at $T\neq 0$ as long as the temperature is relatively small such that our approach is justified.
Indeed, in the weak coupling regime the $T$ dependence of the instanton density is determined
by a simple insertion   $\sim \exp[-(  \frac13 (2N_c+N_f) \pi^2 T^2)\rho^2]$ into the expression
for the density (\ref{instanton}). The temperature dependence also enters the expression for 
$\Delta (T)$. As long as $\Delta (T)$ does not vanish and we are in CS phase, our calculations (\ref{Vinst2})  
are justified, and the critical $\mu_c(T)$ can be estimated as a function of $T$ at
relatively small $T$ as shown in FIG.1.

 \begin{table*}[htb]
\caption{  Results}
\label{table:1}
\newcommand{\m}{\hphantom{$-$}}
\newcommand{\cc}[1]{\multicolumn{1}{c}{#1}}
\renewcommand{\tabcolsep}{2pc} 
\renewcommand{\arraystretch}{1.2} 
\begin{tabular}[c]{|c|c|c|c|}
\hline 
 & $N_c\!=\!3,\; N_f\!=\!2$ & $N_c\!=\!3,\; N_f\!=\!3$ & $N_c\!=\!2,\;
 N_f\!=\!2$ \\ 
\hline 
$\mu_{Bc}/\Lambda$ & 2.3 & 1.4 & 3.5 \\
\hline
$\mu_{Ic}/\Lambda$ & 2.6 & 1.5 & 3.5 \\
\hline
\end{tabular}
\end{table*}
  We should emphasize that in our picture the nature of the phase transition
  is universal and it is not sensitive to the specific values of $N_c$ and $N_f$,
  in spite of the fact that the ground state of the superconducting phase is very sensitive to the  
   values of $N_c$, $N_f$ and quark mass (CFL,  2SC,  crystalline or even more complicated phases).
 
 \begin{figure}[h]
\hspace{-.8cm}
\includegraphics[scale=0.5, clip=true, angle=0,
draft=false]{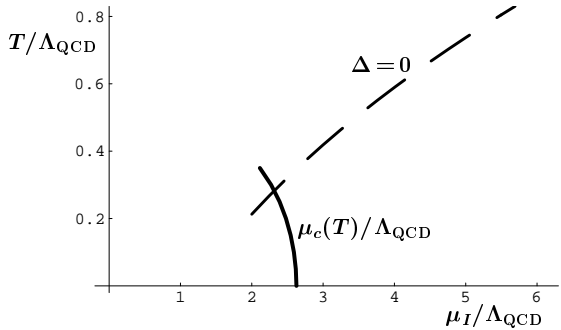}
\caption{\label{fig1} Critical isospin chemical potential for the
  confinement-deconfinement phase transition as a
  function of temperature (solid curve). The dashed curve represents
  the largest temperatures that can be reached in our approach, 
  given by $0.567 \Delta$ (see text for more details).}
\end{figure}
 \section{ Numerical Results.} 
 The critical chemical potential as a function of temperature is
 implicitly given by $a(\mu_c(T),T)=8 f^2(\mu_c(T))/\mu_c^2(T)$.  
We can calculate $a(\mu_c(T),T)$ from (\ref{Vinst2}).  We
 are however limited to temperatures where Cooper pairing takes place,
i.e. for $T\leq0.567 \Delta$ \cite{pr}.
We have determined the critical chemical potential in different cases
  at nonzero baryon or isospin chemical potential.  We
find that the value of the critical chemical potential at $T=0$ are 
given by (we use $m_s\!\simeq 150 MeV $ which is numerically close
to $ 0.75\Lambda$ for $N_f\!=\!3$).
 
As an example, we explicitly show the
results as a function of temperature 
for $N_c=3$ at nonzero $\mu_I$ in FIG.~\ref{fig1}, 
where direct lattice calculation are possible. We notice that with our
conventions the transition from the normal phase to pion condensation
happens at $\mu_I\!=\! m_{\pi}/2$. 

As  expected (see Table 1), for given $N_c$ the critical value for $\mu_c$ decreases when $N_f$ increases.
This is due to the fact that an extra fermion degree of freedom
suppresses  the instantons, such that the instanton density becomes smaller. As direct consequence
of that  suppression  the critical value $\mu_c$ at the point 
when the instanton dilute gas approximation breaks down is smaller for $N_f=3$ than for
 $N_f=2$.
 
  As a final remark:    while we expect that the instanton density
(\ref{Vinst2}) suffers from large uncertainties at $\mu\sim \mu_c$,
 the numerical results for $\mu_c(T)$ are
  not very sensitive to   
  these uncertainties  due to the extraction of the large power  from the instanton density,
$  \Lambda_{QCD}/\mu_c\sim \sqrt[b]{a(\mu_c(T), T)}, ~~b\sim 11/3 N_c-2/3N_f$.

\section{Instanton Quarks as Monopoles. Confinement.}
Having formulated our conjecture and the results which follow from it,
the  question about  the relation 
between    the standard  't~Hooft and Mandelstam \cite{Hooft} picture 
of the  confinement  and  our proposal (when confinement is due to  the 
instanton-quarks)  can be formulated.
The key
point of the 't~Hooft - Mandelstam approach is the assumption that
dynamical monopoles in QCD exist and Bose condense. 
The goal of this section is to  argue
that  the instanton-quarks carry the magnetic charges. 
Therefore, in principle, they may play the role of the dynamical monopoles which are the key players 
  in the 't~Hooft and Mandelstam \cite{Hooft} framework. In this case 
  both pictures could be the two sides of the same coin. 
 
Expression   (\ref{CG2}) clearly shows  that the statistical
ensemble of particles interact according to the Coulomb law. 
An immediate suspicion following
from this observation is that these particles carry a magnetic and/or
electric charge, since charges of that type interact precisely in the
above manner.   This
suspicion will be corroborated in a moment.  
  The charges
$Q_a$ were originally introduced in a very formal manner so that
the QCD effective low energy Lagrangian (\ref{chiral}) can be written in
the dual CG form (\ref{CG2}). In the previous sections we presented arguments that the particles $Q_a$
carry fractional topological charges and can be identified with instanton quarks.
Now we shall argue that these particles also carry the magnetic charges.
 
As a short detour,  let us remind few important
 results regarding the  $SU(2)$ Georgi-Glashow model in the weak
coupling regime, with a $\theta$-term
 when the scalar $\Phi^a$
have a large VEV.  
The monopole solution can be constructed explicitly
and the well-known Witten's effect \cite{Witten1}, where the monopole
acquires an electrical charge, takes place. Let $\cal{N}$ denote the generator of large gauge
transformations corresponding to rotations in the $U(1)$ subgroup of
$SU(2)$ picked out by the gauge field, i.e. rotations in $SU(2)$ about
the axis $n^a=\frac{ \Phi^a}{|\Phi^a|}$.  Rotations by an angle of
$2\pi$ about this axis must yield the identity for arbitrary
configurations, which implies \cite{Witten1} that the magnetic
monopoles carry an electric charge proportional to $\theta$. Indeed,
\bea
\label{7}
1= e^{i2\pi \cal{N}}=e^{i2\pi\frac{Q}{e}- i\theta \frac{e M}{4\pi}},
\eea
where, 
\bea
\label{8}
  M=\frac{1}{v}\int d^3xD_i\Phi^aB_i^a ,~~
   Q=\frac{1}{v}\int d^3xD_i\Phi^aE_i^a, \nn
\eea
are the magnetic and electric charge operators respectively, expressed
in terms of the original fields, and $v$ is the vacuum expectation value $\la\Phi^a\ra$ at
infinity. The combination $\frac{eM}{4\pi}=n_m$ in Eq. (\ref{7}) is an
integer and determines the magnetic charge of the configuration.  As
usual, it is assumed that (\ref{7}) remains correct in the strong
coupling regime when $v$ is not large and/or in the more radical case
when $\Phi^a$ is not present in the original formulation. Indeed, as
explained in \cite{Hooft1} the existence of $\Phi^a$ is not essential
and some effective fields may play its role. 
One finds that monopoles
do exist and the Witten effect expressed by formula (\ref{7}) remains
unaltered even when monopoles appear as singularities in the course of
the gauge fixing procedure as described in \cite{Hooft1}.

Restricting attention to terms which are proportional to the
$\theta$-parameter, a comparison between the CG representation,
Eq.(\ref{CG2}), and Eq.(\ref{7}) will now be carried out. From
the CG representation,
Eq.(\ref{CG2}) the relevant term is the total charge, $Q_{net}$, of the
configuration, while in Eq.(\ref{7}) the relevant factor is the total
magnetic charge $\frac{e M}{4\pi}$ for each time slice.  The following
identification is then made\footnote{  Of course we assume here   that a configuration is
static, or slowly depending on time. Therefore,
the identification (\ref{9}) should be considered as  a relation if the instanton quarks
$Q_a$ were treated as classical sources. It is definitely not the case for the dynamical system
under study.
Nevertheless,  relation (\ref{9})  serves as a good argument suggesting that instanton quarks
carry the magnetic charges. The crucial questions are:  can these monopoles
propagate far away from each other? do these monopoles
   condense?},
\bea
\label{9}
\label{identity}
Q_{net}=\frac{e M}{4\pi}= n_m \in {\rm Z\!\! Z}.
\eea 
From these simple observations one can immediately deduce that our
fractional magnetic charges $Q_a$ cannot be related to any
semi-classical solutions, which can carry only integer charges;
rather, configurations with fractional magnetic charges should have
pure quantum origin.   

One should notice here that the
connection between monopoles and instantons on the classical level is
not a very new idea \cite{nahm}.  Indeed,  for example,  quite recently, such a
relation was established for the periodic instantons (also called
calorons) defined on $R^3\times S^1$  
\cite{vanbaal}, see also \cite{diakonov} and  \cite{khoze} where  
monopoles and instantons  are intimately related objects in semiclassical 
construction.

  Furthermore, a similar relation was seen in the study
of Abelian projection for instantons \cite{Polikarpov,Brower}, albeit
at the classical level.  In particular in ref. \cite{Brower} it was 
demonstrated that the instanton's topological charge, $Q$, is given in
terms of the monopole  charge $M$   forming the  
loop as follows $Q=\frac{eM}{4\pi}$.  This formula is very
similar to our relation (\ref{9}), where the total topological charge,
$Q_{net}$, for a configuration containing a number of particles,
 described by the system (\ref{CG2}) was
identified with the total magnetic charge for each Euclidean time slice for the
same configuration.  Further to
this point, lattice simulations do not contradict this picture where
large instantons induce the magnetic monopole loops forming large
clusters, see e.g.\cite{Diakonov} and references therein. 

We conclude this section with few following remarks.\\
1). The relation between topological charge in 4d and magnetic charge
in 3d is  understood  only on the level of  classical equations of motion, \cite{nahm}-\cite{Brower}.
However, this knowledge does not provide us with answers on the crucial questions 
such as: `` what is dynamical properties  of these monopoles?",  ``do they condense or, rather,
they   propagate  only for  short distances for a short period of time?"\\
2). A similar to eq.(\ref{identity})  identification
 could be made for a different system with large $\mu$ (\ref{CG})
  when only small size  $\rho\sim \mu^{-1} $ instantons are present. However,  in this case
  it is quite obvious that
  the description in terms of the monopole loops makes no sense  because the typical size of 
  the loops is very small, of order $ \sim \mu^{-1} $, and the magnetic charge is obviously
  screened  on large distances. Therefore, monopole charge of constituents
   play no  role for such ensembles. \\
  3). In contrast with the small   instantons,  the constituents of
  large size instantons (instanton quarks with charge
  $1/N_c$ ) may propagate far away from each other. In this case 
  the description in terms of the instanton quarks which carry the monopole charges
   could be appropriate.  For such configurations the  magnetic charges
  of the instanton quarks should manifest themselves in some way. 
  In particular, if the magnetic
charges Bose-condense, this indicates the onset of quark
confinement. To investigate the possibility for such a condensation
an expression for the magnetic charge creation operator, $\MM$, must
be found and its VEV (magnetization) calculated. 
Such a program is   very 
ambitious, and   obviously beyond  the scope of the present work.
\section{Concluding Comments } 
\subsection{Main Results}
The main leitmotiv of this talk is based on the  conjecture that the 
confinement-deconfinement phase transition at nonzero chemical
potential and small temperature   is driven by instantons.
The instantons qualitatively change the shapes at the transition: they small 
  well-localized objects  at large $\mu\gg \mu_c$; they become arbitrary
  large, strongly overlapped configurations at small $\mu\ll \mu_c$ in which case description in terms of the instanton quarks
  become appropriate. 
  Let us emphasize again: the instanton quarks are point like defects
  which have pure quantum origin and can not be described
  as semiclassical configurations.  They are characterized by $4$ translational collective variables,
  such that $k$ units of the topological charge are represented by coherent superposition of
  $N_c$ instanton quarks (per unit charge) to  make together $4N_c k$ collective variables.
   This number precisely matches the 
  number of the instanton parameters with  topological charge $k$.
 While  the instanton quarks can be arbitrary far away from each other, they keep the information
about their origin; they are correlated.
Therefore, instanton quarks form  not a random, but rather, the  coherent large size configurations.

Furthermore we make a
quantitative prediction for the critical value of the chemical
potential where this transition between two descriptions takes
 place: $\mu_c\sim3\Lambda_{\rm QCD}$ at $T=0$.  This prediction can be
readily tested on the lattice at nonzero isospin chemical potential.
\subsection{Future Directions}
There are well established lattice method which allow to 
introduce isospin chemical potential into the system, see
e.g. \cite{kogut}. 
Independently, there are well- established lattice methods which allow to measure
the topological charge density distribution, see e.g.
\cite{Horvath,Gattringer,Horvath1}.
We claim that the topological charge density distribution measured as
a function  of $\mu_I$ 
will experience sharp changes at the same critical value
$\mu_I=\mu_{c}(T)$  
where the phase transition (or rapid crossover) occurs.
Indeed, the changes in the topological charge
density distribution are expected due to the fundamental differences in
$\theta$ dependence in two different regimes. We identify these changes 
with confinement-deconfinement transition based on the arguments presented above.
We strongly advocate the lattice community to perform such an analysis to see
whether corresponding `` accidental coincidence"  indeed takes place.
Such an analysis would provide an unique opportunity  to study a transition
 from ``Higgs -like"  gauge theory to ``Non -Higgs" gauge theory by varying 
the external parameter $\mu_I$ which plays the role of the vacuum expectation value
of a Higgs field\footnote{ ``Higgs -like" 
gauge theories characterized by  some finite expectation value of the Higgs field,
when topological defects have finite size, $\theta$ dependence  is trivial, $\sim \cos \theta$,
and weak coupling regime is realized. This is  in contrast with ``Non -Higgs" gauge theories,
like QCD at zero temperature  and chemical potential when no fundamental scalar fields 
exist, $\theta$ dependence appears in form of 
$\theta/N_c$ and weak coupling regime can not be achieved in description of the large distance physics.}.
In such an analysis one could explicitly study what is happening with finite size
 instantons (which are under complete theoretical control
at large $\mu_I> \mu_c$) when transition from weak coupling regime to strong coupling regime 
occurs.
\subsection {Relation to Other Studies}
Here we would like to make few comments on relation to 
other works. 

{\bf i).}  As we already mentioned, at the intuitive level there seems
 to be a close relation between     
instanton quarks and the ``periodic instanton" 
\cite{vanbaal,diakonov,Gattringer}. 
Indeed, in these papers  it has been shown that the large size
instantons and monopoles are intimately connected
and  instantons have the internal structure resembling the
instanton-quarks.  Also, it has been shown that the
constituents carry the magnetic charges. More than that, 
it has been also argued that large size instantons   likely were missing 
in the lattice simulations, which is consistent with the picture advocated
in the present work.
Unfortunately,
one should not expect to be able to account
for large instantons using semiclassical technique   
to bring this intuitive correspondence onto the quantitative level.
However, such a  mapping  may help us to  understand   the relation 
between pictures advocated by   't~Hooft and Mandelstam \cite{Hooft}
on one hand and picture where  
instanton-quarks are the key players, on the other hand. 

{\bf ii).} There seems
 to be another close relation
 (albeit at the  intuitive level) 
  between   the instanton quarks   and configurations with center vortices and nexuses with fractional fluxes $1/N_c$, see recent papers on the subject and earlier references therein\cite{Engelhardt}. In particular, 
 the total topological charge for entire configuration  in both cases is always integer. Locally, however, essentially independent units carry fractional  charges $1/N_c$. While the geometrical and topological 
 properties are very similar in both cases,
  there is, however,  a fundamental difference between the two: center vortices/nexuses are classical configurations, while the instanton quarks (and everything which accompanying  them) have  pure quantum origin. This remark is also applied  to the ``periodic instanton" mentioned above. 
  This difference, in particular, manifests itself for the gauge group $G$  different from  $SU(N_c)$.
  In this case the fractional topological charge carried by instanton quarks is $1/C_2(G)$,
  see Section III. At the same time,
  in general, $C_2(G)$ is not related to the center of the group playing a crucial role
  in construction of center vortices\cite{Engelhardt}.   
  
{\bf iii).}  Using  the overlap formalism for chiral fermions \cite{Narayanan:1993ss},
 it has been demonstrated  \cite{Edwards:1998dj} that    there is  a strong  
  evidence for there
   existence of gauge field configurations with fractional topological charge $Q=1/2$ for
  $SU(2)$ gauge theory. 
  
{\bf iv).} There is an interesting recent development in lattice computations
 which in principle would allow to study the topological charge fluctuations 
in QCD vacuum without any assumptions or guidance based on some
specific models for  QCD vacuum configurations\cite{Horvath1}.
Our remark here that the picture based on the instanton quarks advocated here is consistent
with these recent lattice results\cite{Horvath1}. 
Indeed, the most profound finding of ref.\cite{Horvath1}
is demonstration that the topological density distribution in QCD has `` inherently global " structure.
It is definitely consistent with our picture when the point like instanton quarks can be 
far away from each other, but still keep the correlation  at arbitrary large distances.

Another interesting observation by ref.\cite{Horvath1} can be explained as follows.
If 4D structures of finite size  (such as  instantons with finite size $\sim \Lambda_{QCD}$) 
dominate the continuum limit, than these coherent regions of size $\sim \Lambda_{QCD}$ 
should exhibit scaling behavior when the lattice spacing is changed. This   feature 
has not been observed in ref.\cite{Horvath1}. Therefore, it has been suggested that, in physical units,
the corresponding 4D structures should shrink to mere points in the continuum limit.
Such an observation is  certainly not in contradiction with our picture where instanton quarks are 
indeed, the effective 4D point like constituents classified by  $4$ translational zero modes.  

As we discussed earlier in the text, the instanton quarks in static limit carry the magnetic charges.
At the same time, the magnetic charge of the entire large-size instanton (with all its constituents
with fractional charges $1/N_c$)
must be zero. Therefore instanton quarks are attached to each other by magnetic strings
such that total magnetic flux of whole system is zero. While the fluxes are $1/N_c$, they can 
be probed by quarks in fundamental representation.
This picture, again, is consistent with feature of the ``{\it sceleton} "
(minimal hard-core substructure exhibiting the global behavior)   from ref.\cite{Horvath1}
 which is viewed as a network
of world lines for point-like objects.

   Finally, the dual picture of our CG representation (describing the instanton quarks) is nothing but the effective chiral  lagrangian for Goldstone fields, see eq.(\ref{chiral}). This `` obvious" connection between confinement and chiral symmetry breaking phenomenon in our framework
  is consistent with speculation   of ref.\cite{Horvath1}
that  the corresponding long distance correlations  might be   associated with long range propagation  of Goldstone fields. 
 
 It is too early to say whether ref.\cite{Horvath1} finds precisely the features we have been advocating
 to exist for quite a while\cite{JZ}, but the results of ref.\cite{Horvath1} look very exciting and promising  to us.

{\bf v).} There seems
 to be that instanton quarks have been identified  in the brane construction in SUSY gauge theories
as $D0$ -brane \cite{Brodie:1998bv}. While these  objects were called as `` fractional instantons" or ``merons" in ref. \cite{Brodie:1998bv}, they obviously have all features of the instanton quarks described above.
In particular,  the  objects    from ref.  \cite{Brodie:1998bv}
 are point like configurations classified by four translational
 collective variables, precisely as discussed above.
  It also has been argued  \cite{Brodie:1998bv}  that they condense in ${\cal N}=1$ SYM 
  which leads to  the confinement in the theory. At the same time, it has been argued
  that the same fractional constituents ($D0$ -branes) do not play any role in 
  ${\cal N} =2 $ 
  Seiber-Witten model \cite{Brodie:1998bv}, see also\cite{Buchel:2001nw}.
  This is perfectly consistent with our proposal that the instanton 
  quarks are not important in ``Higgs- like" gauge theories, but play a crucial role
  in ``Non-Higgs" gauge theories. More than that, we conjecture that the instanton quarks
  is the driving force for the phase transition separating  these two fundamentally 
  different types of gauge theories. We suggest to use  chemical potential $\mu$ 
  as a parameter which allows us to interpolate between these two types of behavior.

{\bf vi).} As the final remark: the $\theta$    parameter played a key role in  all discussions 
presented above.  
However, the region of $\mu_c(T)$ where transition is expected to occur (see Table 1)
is not very sensitive to value of $\theta$. Indeed, the $\theta$ dependence in   physical
observable  comes with extra suppression $\sim m_q$ which is very small factor.
 This is exactly the reason why all results for $\mu_c(T)$
are quoted  for $\theta=0$. This is definitely not the case when transition from normal to superfluid 
phase is considered as a function of baryon  chemical potential at $N_c=2$,
or as a function of isotopical chemical potential at $N_c\geq 3$.
In these cases the transitions are happening at $\mu\sim m_{\pi}(\theta)$
where very nontrivial dependence $\mu_c(T)$ on  $\theta$ is expected\cite{Metlitski:2005db}.

\noindent
 \section*{  Acknowledgements}

\noindent
Author thanks Pierre van Baal for numerous, very  insightful  and never ending 
discussions on the subject.
 Author also thanks all collaborators of the papers \cite{TZ}-\cite{JZ}
which constitute  the main bulk of this talk.  Author also thanks Rajamani Narayanan  for the correspondence regarding papers  \cite{Narayanan:1993ss}
and  \cite{Edwards:1998dj},
Alex Buchel 
for correspondence regarding the papers \cite{Brodie:1998bv} and \cite{Buchel:2001nw},
and Michael Creutz for correspondence regarding the papers
\cite{Metlitski:2005db}.
I am also thankful to the organizers of the Light Cone Meeting, Cairns, Australia, 2005, for inviting  me    
 to speak on this subject.   The work
 was supported, in part, by the Natural Sciences and Engineering
Research Council of Canada.

\end{document}